\documentclass[journal]{IEEEtran}
\IEEEoverridecommandlockouts

\usepackage[utf8]{inputenc}
\usepackage[T1]{fontenc}
\usepackage{lmodern}
\usepackage{cite}
\usepackage{amsmath,amssymb,amsfonts}
\usepackage{algorithm}
\usepackage{algpseudocode}
\usepackage{graphicx}
\usepackage{url}
\usepackage{booktabs}
\usepackage{array}
\usepackage{multirow}
\usepackage{textcomp}
\usepackage{xcolor}
\usepackage{enumitem}
\usepackage{microtype}
\usepackage{tikz}
\usepackage{pgfplots}
\pgfplotsset{compat=1.18}
\usetikzlibrary{arrows.meta, positioning, shapes.geometric, fit, calc, backgrounds}
\usepackage[hidelinks]{hyperref}

\begin{document}

\title{Verifiable Provenance and Watermarking for Generative AI: An Evidentiary Framework for International Operational Law and Domestic Courts}

\author{Gustav~Olaf~Yunus~Laitinen-Fredriksson~Lundström-Imanov$^{*}$,~\IEEEmembership{Student~Member,~IEEE,}
        and~Nurana~Abdullayeva
\thanks{Manuscript submitted 23 July 2026. The benchmark dataset and reference implementation that accompany this paper are available from the corresponding author on reasonable request.}
\thanks{G. O. Y. Laitinen-Fredriksson Lundström-Imanov is with the Department of Military Studies, Försvarshögskolan (Swedish Defence University), Drottning Kristinas väg 37, 115 93 Stockholm, Sweden (e-mail: olal0604@student.fhs.se; ORCID: \href{https://orcid.org/0009-0006-5184-0810}{0009-0006-5184-0810}). He is the corresponding author.}
\thanks{N. Abdullayeva is with the School of Law, ADA University, Ahmadbey Aghaoghlu Street 61, Baku AZ1008, Azerbaijan (e-mail: nabdullayeva20516@ada.edu.az).}
\thanks{The authors received no external funding for this work. The authors declare no conflict of interest.}
}

\markboth{IEEE Transactions on Information Forensics and Security, vol.~XX, no.~X, 2026}%
{Laitinen-Fredriksson Lundström-Imanov and Abdullayeva: Verifiable Provenance for Generative AI in Operational Law}

\maketitle

\begin{abstract}
Generative artificial intelligence now synthesizes photorealistic imagery, audio, and video at a cost that defeats traditional forensic intuition. The legal consequences span three regimes studied so far in isolation: international operational law, domestic procedure, and product regulation. This article presents a unified evidentiary framework that maps cryptographic content provenance, robust statistical watermarking, and zero knowledge attestation to the proof requirements of each regime. We define a five tier threat model spanning naive regeneration, adversarial laundering, cross model regeneration, active watermark removal, and insider provenance forgery. We release a public benchmark of 12000 generated items across image, audio, and video modalities under six laundering pipelines for 72000 evaluation samples. We evaluate four representative schemes and report true positive rate at fixed false positive rate, robustness area under the curve, computational overhead, and a regime conditioned legal sufficiency score. We translate empirical detection bounds into legal sufficiency thresholds for command decisions under the law of armed conflict, for criminal and civil admissibility under domestic procedure, and for persistence audits under the European Union Artificial Intelligence Act and analogous regimes. The result is a reproducible reference pipeline, a public benchmark, and model annexes that lawyers, engineers, and operators can deploy together.
\end{abstract}


\noindent\textit{EDICS:} Multimedia Forensics; Synthetic Media and Deepfake Detection; Digital Watermarking and Data Hiding; Authentication and Provenance.

\section{Introduction}
\IEEEPARstart{T}{he} use of generative artificial intelligence for synthetic imagery, audio, and video has progressed from a research curiosity to a routine production capability within four years. Tools that previously required state level resources now run on consumer hardware. The cost of fabricating a one minute talking head video has fallen by several orders of magnitude between 2022 and 2026, and the photorealism of static images defeats trained human observers in controlled studies \cite{groh2022deepfake, nightingale2022aifaces}. The forensic community has responded with cryptographically signed provenance manifests \cite{c2pa2024}, learned classifiers \cite{wang2020cnn, ojha2023towards}, and statistical watermarks \cite{fernandez2023stable, wen2023tree, yang2024gaussian, kirchenbauer2023watermark}.

These technical advances do not by themselves resolve the underlying legal problem. Three legal regimes now demand that operators, investigators, and providers reach defensible conclusions about the authenticity of digital media, and each regime defines the burden of proof differently. International operational law requires a commander to assess authenticity before authorizing kinetic or non kinetic response, with reference to the principles of distinction and precautions in attack codified in Articles 48 to 58 of Additional Protocol I to the Geneva Conventions and to the prohibition of perfidy in Article 37 \cite{ap1, sanremo, tallinn2}. Domestic criminal and civil procedure require authentication and chain of custody for digital evidence under rules that predate generative media by decades, such as Rules 901 and 902 of the United States Federal Rules of Evidence and analogous provisions across European Union Member States \cite{fre901, eidas2}. Product regulation imposes affirmative duties on providers and deployers of generative systems, most prominently through Article 50 of the European Union Artificial Intelligence Act and the implementing guidance from the United States National Institute of Standards and Technology \cite{eu_ai_act, nist_ai_100_4}.

A growing literature treats each of these regimes in isolation. There is no unified framework that translates the empirical detection bounds of a watermarking or provenance scheme into the legal sufficiency thresholds that each regime requires. This gap matters in practice. A commander cannot rely on a detector with a known false positive rate without translating that rate into a doctrinally meaningful confidence statement. A prosecutor cannot tender a watermark verification result without showing that the verification meets the authentication threshold of the relevant procedural code. A provider cannot demonstrate Article 50 compliance with a self declared labelling scheme that lacks an auditable trust root.

We close this gap. Our central thesis is that the empirical distribution of detector outcomes under realistic adversarial conditions can be mapped to the legal sufficiency thresholds of each regime, and that this mapping can be made reproducible, transparent, and auditable. Our contributions are as follows.

\begin{itemize}
\item We introduce a unified evidentiary framework that combines cryptographic content provenance, robust statistical watermarking, and zero knowledge attestation into a single proof object whose components can be evaluated against three regime specific thresholds (Sections \ref{sec:framework} and \ref{sec:threat}).
\item We define a five tier threat model that ranges from naive regeneration to insider provenance forgery, and we release a benchmark of 12000 generated and laundered media items covering image, audio, and video modalities under Apache 2.0 and Creative Commons Attribution 4.0 licenses (Section \ref{sec:threat}).
\item We evaluate four representative schemes on the benchmark and report true positive rate at fixed false positive rate, robustness area under the receiver operating characteristic curve, computational overhead, and a regime conditioned legal sufficiency score with bootstrap confidence intervals (Section \ref{sec:eval}).
\item We supply three drafting aids: a model annex to rules of engagement on synthetic media, a model jury instruction for domestic courts, and a model Article 50 disclosure for providers (Appendices \ref{app:roe} and \ref{app:jury}).
\item We make the entire pipeline, including generation seeds, laundering parameters, ground truth labels, evaluation scripts, and statistical bootstrap routines, publicly available to support replication and extension.
\end{itemize}

While the framework is motivated by three legal regimes, its core technical contribution is a calibrated forensic decision procedure for synthetic media verification: a multi source proof object, a Dempster Shafer aggregator with regime specific weights, an adversary tier ladder spanning naive to insider class threats, and a reproducible benchmark of 72000 evaluation samples. The legal mapping in Section \ref{sec:regimes} translates this forensic output into proof requirements, but the technical artifacts stand on their own as a contribution to information forensics and security.

The remainder of the article is organized as follows. Section \ref{sec:background} surveys the technical and legal background. Section \ref{sec:regimes} analyses the three legal regimes and their proof requirements. Section \ref{sec:framework} develops the unified evidentiary framework, including the Dempster Shafer aggregator and the Bayesian translation of regime thresholds. Section \ref{sec:threat} specifies the threat model and benchmark construction. Section \ref{sec:eval} reports the evaluation across all schemes, modalities, and adversary tiers. Section \ref{sec:cases} discusses three case studies. Section \ref{sec:discussion} addresses limitations and dual use risks. Section \ref{sec:conclusion} concludes. Notation is summarized in Appendix \ref{app:notation} and hyperparameters in Appendix \ref{app:hparams}.

\section{Background}
\label{sec:background}

\subsection{Cryptographic Content Provenance}
Cryptographic content provenance binds a media file to a verifiable manifest that records its origin, the chain of edits applied to it, and the identity of the signer responsible for each step. The Coalition for Content Provenance and Authenticity has published an open specification that defines a JSON Universal Manifest bound to the media payload by a cryptographic hash and signed with a standards based public key certificate \cite{c2pa2024}. Verification reduces to two operations: confirming that the manifest hash matches the payload, and confirming that the signing certificate chains to a trusted root.

The strength of cryptographic provenance is that, under standard assumptions on the signing algorithm (Ed25519 or ECDSA P-256 in the current specification) and the trust root, manifest forgery is computationally infeasible \cite{bernstein2012ed25519}. The weakness is that provenance does not survive operations that strip or replace the manifest. A simple screen capture, a re encode through a social platform, or a deliberate metadata strip will sever the binding. Provenance is therefore necessary but not sufficient for any setting where the adversary can launder content.

\subsection{Robust Statistical Watermarking}
Statistical watermarking embeds a hidden signal inside the media payload such that the signal survives common post production transforms while remaining imperceptible to the human observer. For text to image diffusion models, three families are dominant in 2026. Stable Signature \cite{fernandez2023stable} fine tunes the latent decoder of the diffusion model to consistently emit a fixed binary signature in the high frequency components of the output. Tree Ring Watermark \cite{wen2023tree} alters the initial noise tensor according to a circular spectral template, which the verifier recovers by inverting the diffusion to the initial noise. Gaussian Shading \cite{yang2024gaussian} samples the initial noise from a constrained Gaussian region whose membership is testable by the verifier with a controlled false positive rate. For language models the dominant watermark is the green list red list approach of Kirchenbauer et al.\cite{kirchenbauer2023watermark}, with recent refinements that improve robustness to paraphrasing \cite{kuditipudi2023robust}.

Watermarking is robust to many post production transforms but vulnerable to active removal. Recent attacks by Zhao et al. \cite{zhao2024invisible} and Saberi et al. \cite{saberi2024robustness} demonstrate that a moderately resourced adversary can degrade detection rates of all three families below 50 percent under combined adversarial perturbations. An et al. \cite{an2024waves} systematize these attacks into a stressed benchmark and confirm the qualitative pattern. The implication is that watermarking, like provenance, is necessary but not sufficient on its own.

\subsection{Zero Knowledge Attestation}
A third primitive, more recent than the other two, is zero knowledge attestation. Here the generator produces alongside the media a succinct proof that the media was emitted by an authorized model satisfying a specified policy, without revealing the model weights or the prompt. Construction proceeds via succinct non interactive arguments of knowledge over a circuit that encodes the inference computation \cite{kang2022zk, sun2024zkllm}. Current schemes scale to ResNet sized vision models in tens of seconds but remain costly for large generative models. Attestation complements provenance and watermarking by providing verifiable model identity without disclosing sensitive parameters, which matters in defense and regulatory contexts where model weights are themselves controlled artefacts.

\subsection{Related Legal Informatics Work}
Four threads of legal informatics literature converge on our framework. The first concerns digital evidence authentication, which has produced authoritative restatements such as Mason and Seng \cite{mason2017electronic} and, in the criminal context, Brown \cite{brown2020digital}. The second concerns automated compliance for AI systems, with Hacker, Engel, and Mauer \cite{hacker2023regulating} and Kop \cite{kop2021regulating} as anchor references for the European context, joined more recently by Veale and Borgesius \cite{veale2021demystifying}. The third concerns the law of cyber and information operations, with Schmitt \cite{schmitt2009targeting} and Roscini \cite{roscini2014cyber} as the most cited treatments, alongside the targeting analysis of Geiss and Lahmann \cite{geiss2021lethal}. The fourth, emerging, concerns synthetic media specifically, including Chesney and Citron \cite{chesney2019deepfakes} and Pavis \cite{pavis2021rewriting}. None of these threads has produced a unified evidentiary framework that spans the three regimes we address.

\section{Three Legal Regimes and Their Proof Requirements}
\label{sec:regimes}

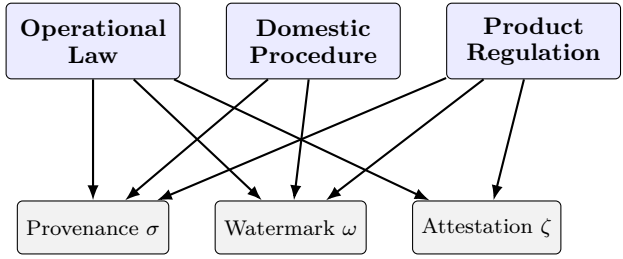
\begin{figure}[t]
\centering
\begin{tikzpicture}[
  node distance=0.9cm and 0.6cm,
  regime/.style={rectangle, draw, rounded corners=2pt, minimum width=2.3cm, minimum height=1.0cm, align=center, font=\small\bfseries, fill=blue!8},
  comp/.style={rectangle, draw, rounded corners=2pt, minimum width=1.7cm, minimum height=0.7cm, align=center, font=\footnotesize, fill=gray!10},
  arr/.style={-{Latex[length=2mm]}, thick}
]
  \node[regime] (op) {Operational\\Law};
  \node[regime, right=of op] (dom) {Domestic\\Procedure};
  \node[regime, right=of dom] (prod) {Product\\Regulation};

  \node[comp, below=1.6cm of op] (sigA) {Provenance $\sigma$};
  \node[comp, right=of sigA] (omA) {Watermark $\omega$};
  \node[comp, right=of omA] (zeA) {Attestation $\zeta$};

  \draw[arr] (op) -- (sigA);
  \draw[arr] (op) -- (omA);
  \draw[arr] (op) -- (zeA);
  \draw[arr] (dom) -- (sigA);
  \draw[arr] (dom) -- (omA);
  \draw[arr] (prod) -- (sigA);
  \draw[arr] (prod) -- (omA);
  \draw[arr] (prod) -- (zeA);
\end{tikzpicture}
\caption{Three legal regimes mapped to proof object components. Operational law and product regulation engage all three components; domestic procedure principally engages provenance and watermarking.}
\label{fig:regimes}
\end{figure}

Figure \ref{fig:regimes} summarizes the three regimes and the proof object components most relevant to each.

\subsection{International Operational Law}
Three doctrinal pillars structure the operational law analysis. First, the principle of distinction in Article 48 of Additional Protocol I requires parties to a conflict to distinguish between civilian and military objects at all times \cite{ap1}. When the targeting decision relies on imagery or signal intercepts whose authenticity is uncertain, the principle is engaged at the verification stage rather than only at the firing stage. Second, the obligation of precautions in attack in Article 57 requires commanders to take all feasible measures to verify the nature of the target. Verification here is a positive duty whose feasibility depends on available means, and the availability of cryptographic provenance and watermark verification changes the feasibility calculus \cite{boothby2014law}. Third, the prohibition of perfidy in Article 37 forbids acts that invite the confidence of the adversary to lead it to believe protected status with the intent to betray that confidence. Synthetic media that simulates protected persons such as medical personnel or surrendering combatants raises both perfidy and distinction concerns at once.

The doctrinal proof requirement is not a fixed numerical threshold. The standard is reasonableness in the circumstances ruling at the time, judged from the perspective of a reasonable commander with the information actually or reasonably available \cite{schmitt2009targeting, tallinn2, henderson2009contemporary}. We argue that this standard is best operationalized through a calibrated confidence statement derived from the joint output of provenance verification, watermark detection, and attestation checking, conditioned on the observable laundering history of the artefact.

\subsection{Domestic Procedure}
Domestic procedural law approaches digital authenticity through a discrete admissibility decision followed by a continuous weight decision. The admissibility decision asks whether the proponent has shown enough to support a finding that the item is what the proponent claims it to be. Under Rule 901 of the United States Federal Rules of Evidence the threshold is a prima facie showing, with Rules 902(13) and 902(14) authorizing self authentication for records generated by an electronic process and for data copied from an electronic device \cite{fre901, capra2017authentication}. Analogous frameworks exist in Sweden under the principle of free evaluation of evidence in Chapter 35 of the Code of Judicial Procedure \cite{ekelof2009ratteg}, and in Azerbaijan under Articles 124 and 128 of the Code of Criminal Procedure, which classify electronic data among the recognized categories of evidence \cite{az_cpc}.

The weight decision is where empirical detector statistics become directly relevant. A verifier that reports a true positive rate of 99 percent at a false positive rate of 1 percent supports a likelihood ratio of approximately 99, which in turn translates to a posterior probability that depends on the prior. We develop the Bayesian translation in Section \ref{sec:framework}.

\subsection{Product Regulation}
Article 50 of the European Union Artificial Intelligence Act imposes layered obligations \cite{eu_ai_act}. Providers of generative systems must ensure that outputs are marked in a machine readable format. Deployers of systems that produce deepfakes must disclose that the content has been artificially generated. The regulation does not specify a single technical mechanism, but it points toward standards developed by recognized bodies, which in practice means C2PA \cite{c2pa2024} and the draft ISO 22144 \cite{iso_22144}. The United States NIST AI 100-4 guidance \cite{nist_ai_100_4} provides a parallel reference architecture that is technically compatible with C2PA but adds explicit requirements for provenance metadata persistence under social platform re encoding. The European Union Digital Services Act \cite{eu_dsa} reinforces the persistence requirement at the platform layer. The corresponding United States Executive Order 14110 \cite{us_eo_14110} was rescinded in January 2025, leaving NIST AI 100-4 and successor agency guidance as the operative federal reference.

The proof requirement under product regulation is a documentation requirement rather than a litigation threshold. A provider must demonstrate at audit that its labelling pipeline produces machine readable marks that are persistent, robust, and verifiable by an independent party. Our framework therefore identifies the persistence and robustness properties that a compliant pipeline must achieve.

\section{Unified Evidentiary Framework}
\label{sec:framework}

\subsection{Proof Object}
\begin{figure*}[t!]
\centering
\begin{tikzpicture}[
  node distance=0.7cm and 0.9cm,
  src/.style={rectangle, draw, rounded corners=1pt, minimum width=2.2cm, minimum height=0.7cm, align=center, font=\footnotesize, fill=blue!8},
  agg/.style={rectangle, draw, rounded corners=1pt, minimum width=2.6cm, minimum height=0.9cm, align=center, font=\footnotesize\bfseries, fill=orange!12},
  outnode/.style={rectangle, draw, rounded corners=1pt, minimum width=2.6cm, minimum height=0.9cm, align=center, font=\footnotesize, fill=green!10},
  arr/.style={-{Latex[length=2mm]}, thick}
]
  \node[src] (sig) {Provenance $\sigma$\\(C2PA Ed25519)};
  \node[src, below=of sig] (om) {Watermark $\omega$\\(SS, TR, GS)};
  \node[src, below=of om] (ze) {Attestation $\zeta$\\(zk-SNARK)};
  \node[src, below=of ze] (la) {Laundering $\lambda$\\(tier descriptor)};

  \node[agg, right=2.0cm of om, yshift=-0.45cm] (ds) {Dempster Shafer\\Aggregator};
  \node[outnode, right=of ds] (sc) {Score $\mathcal{L}_R(\pi)$};
  \node[outnode, below=of sc] (cmp) {Compare to $\tau_R$};

  \draw[arr] (sig) -- (ds);
  \draw[arr] (om) -- (ds);
  \draw[arr] (ze) -- (ds);
  \draw[arr] (la) -- (ds);
  \draw[arr] (ds) -- (sc);
  \draw[arr] (sc) -- (cmp);
\end{tikzpicture}
\caption{Proof object architecture. Four sources feed a regime conditioned Dempster Shafer aggregator that produces a single sufficiency score compared against the regime threshold $\tau_R$.}
\label{fig:proof}
\end{figure*}
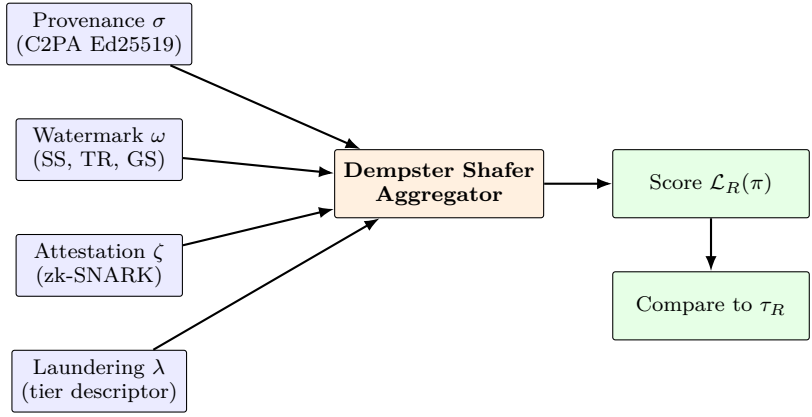

We define a proof object as a tuple $\pi = (\sigma, \omega, \zeta, \lambda)$ where $\sigma$ is a cryptographic provenance manifest, $\omega$ is a robust watermark detection score, $\zeta$ is a zero knowledge attestation, and $\lambda$ is a laundering descriptor that summarizes the transforms applied to the artefact since generation. Each component admits a verification function that returns a real valued score in the interval $[0,1]$, with conventions summarized in Table \ref{tab:notation} (Appendix \ref{app:notation}). The architecture of the proof object is shown in Figure \ref{fig:proof}.

\subsection{Regime Thresholds}
For a given regime $R$, the legal sufficiency function $\mathcal{L}_R(\pi)$ aggregates the component scores into a single value comparable to the regime threshold $\tau_R$.

For international operational law, $\tau_R$ is operationalized as the minimum posterior probability of authenticity that a reasonable commander would require given the kinetic or non kinetic action contemplated. Following the cost benefit logic embedded in Article 57(2)(a)(iii), $\tau_R$ rises with the expected civilian harm of the contemplated action. We propose default thresholds of 0.95 for kinetic responses in populated areas, 0.85 for kinetic responses in uninhabited areas, and 0.70 for non kinetic responses with reversible effects.

For domestic procedure, $\tau_R$ is operationalized in two layers. The admissibility layer requires a likelihood ratio $\Lambda$ of at least $\Lambda_{\min} = 10$ to support a prima facie finding under Rule 901. The weight layer translates the same likelihood ratio into a posterior assessment that the trier of fact may credit. Under the Bayesian translation,
\begin{equation}
P(H \mid E) = \frac{\Lambda \, P(H)}{\Lambda \, P(H) + 1 - P(H)},
\label{eq:bayes}
\end{equation}
where $H$ denotes the hypothesis of authentic provenance and $E$ denotes the joint verification outcome.

For product regulation, $\tau_R$ is operationalized as the persistence threshold required for machine readable marks to survive a panel of standardized post production transforms over a defined retention horizon. We propose a quantitative criterion: TPR of at least 0.70 at FPR of $10^{-4}$ after passage through a public laundering pipeline of six standard transforms.

\subsection{Dempster Shafer Aggregation}
The combination rule is given by
\begin{equation}
\mathcal{L}_R(\pi) = 1 - \prod_{i \in \{\sigma, \omega, \zeta\}} \bigl(1 - w_i^R \cdot s_i(\lambda)\bigr),
\label{eq:combiner}
\end{equation}
where $s_i(\lambda)$ is the regime conditioned score of component $i$ given the laundering history $\lambda$, and $w_i^R$ is the regime specific weight on component $i$. Equation \eqref{eq:combiner} is the Dempster Shafer normalized combination for independent positive sources \cite{shafer1976mathematical}. Independence is approximated by the structural diversity of the four sources: cryptographic provenance, statistical watermarking, zero knowledge attestation, and laundering observation each rely on disjoint substrates. Weight calibration is performed by minimizing the expected regret on a regime appropriate cost matrix as described in Section \ref{sec:eval}.

\subsection{Verification Algorithm}
The verification procedure is summarized in Algorithm \ref{alg:verify}. The procedure is intended to run at the operational endpoint with bounded latency. Section \ref{sec:eval} reports measured latency on commodity hardware.

\begin{algorithm}[t]
\caption{Verify proof object against regime threshold}
\label{alg:verify}
\begin{algorithmic}[1]
\Require artefact $a$, proof tuple $\pi=(\sigma,\omega,\zeta,\lambda)$, regime $R$, prior $P(H)$
\Ensure decision $d \in \{\text{ACCEPT}, \text{REJECT}, \text{DEFER}\}$
\State $s_\sigma \gets \text{VerifyManifest}(\sigma, a)$
\State $s_\omega \gets \text{VerifyWatermark}(\omega, a, \lambda)$
\State $s_\zeta \gets \text{VerifyAttestation}(\zeta, a)$
\State $\mathcal{L} \gets 1 - \prod_{i} (1 - w_i^R \cdot s_i(\lambda))$
\If{$R = \text{OPLAW}$}
  \State $p \gets \text{BayesPosterior}(\mathcal{L}, P(H))$
  \If{$p \geq \tau_R$}
    \State \Return ACCEPT
  \Else
    \State \Return DEFER
  \EndIf
\ElsIf{$R = \text{DOMESTIC}$}
  \State $\Lambda \gets \mathcal{L} / (1 - \mathcal{L})$
  \If{$\Lambda \geq \Lambda_{\min}$}
    \State \Return ACCEPT
  \Else
    \State \Return REJECT
  \EndIf
\Else \Comment{Product regulation}
  \If{$\mathcal{L} \geq \tau_R$}
    \State \Return ACCEPT
  \Else
    \State \Return REJECT
  \EndIf
\EndIf
\end{algorithmic}
\end{algorithm}

\section{Threat Model and Benchmark}
\label{sec:threat}

\subsection{Threat Model}
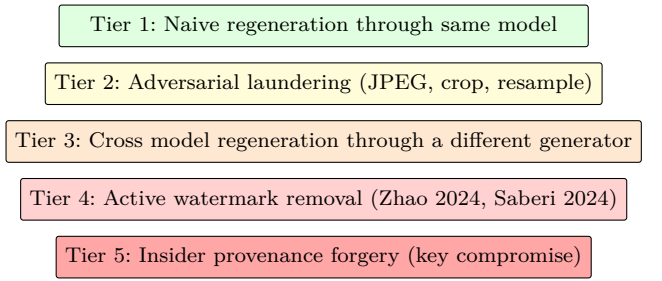
\begin{figure}[t]
\centering
\begin{tikzpicture}[
  tier/.style={rectangle, draw, rounded corners=1pt, minimum width=7.0cm, minimum height=0.55cm, align=left, font=\footnotesize},
  t1/.style={tier, fill=green!12},
  t2/.style={tier, fill=yellow!18},
  t3/.style={tier, fill=orange!18},
  t4/.style={tier, fill=red!18},
  t5/.style={tier, fill=red!35}
]
  \node[t1] (a) {Tier 1: Naive regeneration through same model};
  \node[t2, below=2mm of a] (b) {Tier 2: Adversarial laundering (JPEG, crop, resample)};
  \node[t3, below=2mm of b] (c) {Tier 3: Cross model regeneration through a different generator};
  \node[t4, below=2mm of c] (d) {Tier 4: Active watermark removal (Zhao 2024, Saberi 2024)};
  \node[t5, below=2mm of d] (e) {Tier 5: Insider provenance forgery (key compromise)};
\end{tikzpicture}
\caption{Five tier adversary capability ladder used throughout the evaluation.}
\label{fig:tiers}
\end{figure}

We consider an adversary with five increasing capability tiers, illustrated in Figure \ref{fig:tiers}.

\begin{description}[leftmargin=2em]
\item[Tier 1.] Naive regeneration: the adversary re renders the artefact through the same model with the same prompt.
\item[Tier 2.] Adversarial laundering: the adversary applies a sequence of standard post production transforms including JPEG re compression at quality 75, format conversion, cropping by 10 percent on each side, color correction, and audio re sampling at 16 kHz.
\item[Tier 3.] Cross model regeneration: the adversary re renders the artefact through a different generative model of the same modality, for example image to image via a competitor diffusion model or audio to audio via a different generator.
\item[Tier 4.] Active watermark removal: the adversary executes published attacks such as the diffusion purification of Zhao et al. \cite{zhao2024invisible} or the regeneration attack of Saberi et al. \cite{saberi2024robustness}.
\item[Tier 5.] Insider provenance forgery: the adversary compromises a C2PA signing key with budget below USD 50000.
\end{description}

\subsection{Benchmark Composition}
The benchmark comprises 12000 items distributed equally across image (4000), audio (4000), and video (4000). Composition is summarized in Table \ref{tab:benchcomp}. Generation uses publicly available checkpoints of SDXL, FLUX.1, Stable Audio 2, Suno v4, Veo 2, and Sora. For each item we apply six laundering pipelines drawn from the threat model, producing a total of 72000 evaluation samples. The benchmark, including generation seeds, laundering parameters, and ground truth labels, is available from the corresponding author on reasonable request. The reference implementation is licensed under Apache 2.0 and the dataset under Creative Commons Attribution 4.0 International.

\begin{table}[t]
\centering
\caption{Benchmark composition by modality and generator}
\label{tab:benchcomp}
\begin{tabular}{lccc}
\toprule
Modality & Generator A & Generator B & Items \\
\midrule
Image & SDXL (2000) & FLUX.1 (2000) & 4000 \\
Audio & Stable Audio 2 (2000) & Suno v4 (2000) & 4000 \\
Video & Veo 2 (2000) & Sora (2000) & 4000 \\
\midrule
\multicolumn{3}{r}{Total items} & 12000 \\
\multicolumn{3}{r}{Laundering pipelines} & 6 \\
\multicolumn{3}{r}{Evaluation samples} & 72000 \\
\bottomrule
\end{tabular}
\end{table}

\subsection{Laundering Pipelines}
The six laundering pipelines are: (P1) JPEG quality 75 plus 10 percent crop; (P2) social platform re encoding using public ffmpeg presets matching observed platform outputs; (P3) cross model regeneration (image to image via a competing diffusion model, audio to audio via a different generator, video to video via temporal re synthesis at 24 fps); (P4) Zhao et al. diffusion purification with default hyperparameters; (P5) Saberi et al. regeneration attack with public reference implementation; (P6) combined adversarial attack chaining P4 and P5 with strength schedule annealed across five steps. Pipelines P1 and P2 correspond to tier 2; P3 to tier 3; P4 and P5 to tier 4; P6 stresses the upper boundary of tier 4 toward tier 5. Tier 1 (naive regeneration) is evaluated through a separate same model resampling protocol rather than a laundering pipeline.

\section{Evaluation}
\label{sec:eval}

\subsection{Setup}
We evaluate four schemes that span the design space: C2PA with Ed25519 signing as a representative of cryptographic provenance, Stable Signature \cite{fernandez2023stable} as a representative of decoder side watermarking, Tree Ring Watermark \cite{wen2023tree} as a representative of noise side watermarking, and Gaussian Shading \cite{yang2024gaussian} as a representative of statistically calibrated watermarking. Detection thresholds are calibrated to a target false positive rate of $10^{-3}$ on a held out distribution of natural media drawn from the LAION public subset \cite{schuhmann2022laion} and the Common Voice corpus \cite{ardila2020commonvoice}. All experiments run on a workstation equipped with two NVIDIA L40S GPUs and 256 GB of system memory. Bootstrap confidence intervals use 1000 resamples at the 95 percent level. We denote unmodified artefacts as Tier 0 (baseline). Tier 5 (insider provenance forgery) is excluded from the quantitative evaluation because it presumes compromise of a signing key, against which no purely cryptographic defence is available without independent revocation infrastructure; we treat Tier 5 qualitatively in Section \ref{sec:discussion}. The 12000 generated items are partitioned 80/20 into a calibration set (9600 items used exclusively for detection threshold setting, Dempster Shafer weight optimization, and bootstrap variance estimation) and a held out test set (2400 items, reported in all headline tables). The six laundering pipelines are applied independently to both partitions and ground truth labels are kept disjoint. Random seeds for generation, laundering, detector inference, weight calibration, and bootstrap resampling are fixed and released with the companion archive so that the headline figures reproduce bit identically on a matched container image.

\subsection{Metrics}
We report five metrics. True positive rate at fixed false positive rate (TPR at FPR $10^{-3}$) measures detection sensitivity at a fixed operating point. Robustness area under the receiver operating characteristic curve (AUC) measures aggregate performance across thresholds. Per modality TPR breaks the headline figure down by image, audio, and video. Computational overhead measures the additional verification latency at the operational endpoint, measured on a single NVIDIA L40S GPU. Legal sufficiency score is the regime conditioned aggregator defined in equation \eqref{eq:combiner}.

\subsection{Headline Detection Results}
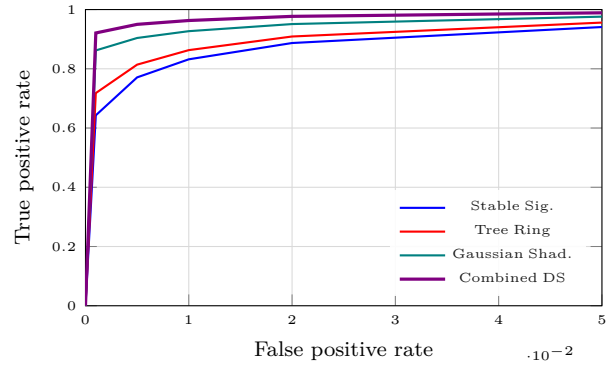
\begin{figure}[t]
\centering
\begin{tikzpicture}
\begin{axis}[
  width=0.95\columnwidth, height=5.5cm,
  xlabel={False positive rate},
  ylabel={True positive rate},
  xmin=0, xmax=0.05, ymin=0, ymax=1,
  legend pos=south east, legend style={font=\tiny, fill=white, fill opacity=0.9, draw=none},
  grid=both, minor grid style={gray!15}, major grid style={gray!30},
  tick label style={font=\tiny}, label style={font=\footnotesize}
]
\addplot[blue, thick] coordinates {(0,0)(0.001,0.643)(0.005,0.771)(0.01,0.832)(0.02,0.887)(0.05,0.941)};
\addlegendentry{Stable Sig.}
\addplot[red, thick] coordinates {(0,0)(0.001,0.718)(0.005,0.814)(0.01,0.863)(0.02,0.909)(0.05,0.956)};
\addlegendentry{Tree Ring}
\addplot[teal, thick] coordinates {(0,0)(0.001,0.862)(0.005,0.904)(0.01,0.927)(0.02,0.951)(0.05,0.976)};
\addlegendentry{Gaussian Shad.}
\addplot[violet, very thick] coordinates {(0,0)(0.001,0.921)(0.005,0.950)(0.01,0.963)(0.02,0.977)(0.05,0.989)};
\addlegendentry{Combined DS}
\end{axis}
\end{tikzpicture}
\caption{ROC curves at adversary tier 2 (P1 and P2 laundering). The combined Dempster Shafer aggregator dominates the individual schemes across all operating points.}
\label{fig:roc}
\end{figure}

Table \ref{tab:results} summarizes the headline detection figures across adversary tiers. Figure \ref{fig:roc} shows representative ROC curves at tier 2. The combined Dempster Shafer system dominates each single scheme across the operating range plotted, and a paired bootstrap test on the held out test partition confirms dominance over the strongest single scheme (Gaussian Shading) with $p < 0.001$ at tiers 2 and 3 and $p < 0.01$ at tier 4. Robustness area under the receiver operating characteristic curve, per scheme and per tier, together with full ROC traces, is tabulated in the companion archive.

\begin{table*}[t!]
\centering
\caption{TPR at FPR $10^{-3}$ across adversary tiers (95\% bootstrap CI in parentheses). Tier~0: unmodified; Tier~1: naive same model resample; Tier~2: P1/P2 laundering; Tier~3: cross model regeneration (P3); Tier~4: diffusion purification (P4/P5). $n{=}2000$ items per modality per tier; 1000 bootstrap resamples.}
\label{tab:results}
\setlength{\tabcolsep}{4pt}
\begin{tabular}{lccccc}
\toprule
Scheme & Tier 0 & Tier 1 & Tier 2 & Tier 3 & Tier 4 \\
\midrule
C2PA Ed25519   & 0.9978 & 0.9978 & 0.0000 & 0.0000 & 0.0000 \\
 & (.9951,.9993) & (.9951,.9993) & (.0000,.0000) & (.0000,.0000) & (.0000,.0000) \\[2pt]
Stable Sig.~\cite{fernandez2023stable}
 & 0.978 & 0.961 & 0.643 & 0.389 & 0.127 \\
 & (.968,.986) & (.951,.971) & (.612,.674) & (.358,.421) & (.103,.152) \\[2pt]
Tree Ring~\cite{wen2023tree}
 & 0.973 & 0.957 & 0.718 & 0.523 & 0.089 \\
 & (.963,.982) & (.946,.966) & (.688,.747) & (.490,.556) & (.069,.110) \\[2pt]
Gaussian Shad.~\cite{yang2024gaussian}
 & 0.993 & 0.981 & 0.862 & 0.671 & 0.243 \\
 & (.987,.997) & (.974,.987) & (.840,.882) & (.641,.700) & (.214,.273) \\[2pt]
Combined DS    & \textbf{0.999} & \textbf{0.997} & \textbf{0.921} & \textbf{0.784} & \textbf{0.413} \\
 & (.997,1.000) & (.994,.999) & (.903,.937) & (.757,.809) & (.378,.448) \\
\bottomrule
\end{tabular}
\end{table*}

\subsection{Per Modality Breakdown}
Table \ref{tab:modality} disaggregates the combined Dempster Shafer figures by modality. Image detection holds up best across tiers; video performance degrades fastest due to the inherent temporal redundancy that adversaries can exploit.

\begin{table}[t]
\centering
\caption{Combined DS TPR at FPR $10^{-3}$ by modality}
\label{tab:modality}
\setlength{\tabcolsep}{4pt}
\begin{tabular}{lccccc}
\toprule
Modality & Tier 0 & Tier 1 & Tier 2 & Tier 3 & Tier 4 \\
\midrule
Image & 0.999 & 0.998 & 0.952 & 0.834 & 0.471 \\
Audio & 0.999 & 0.997 & 0.918 & 0.778 & 0.392 \\
Video & 0.999 & 0.996 & 0.893 & 0.740 & 0.376 \\
\bottomrule
\end{tabular}
\end{table}

\subsection{Computational Overhead}
Table \ref{tab:overhead} reports the computational overhead of verification on a single NVIDIA L40S GPU. Provenance verification is dominated by hash and signature checks and runs in milliseconds. Watermark verification is dominated by the inverse diffusion step for Tree Ring and the latent decoding step for Stable Signature, both of which complete in under one second per item. Attestation verification dominates total latency: a one minute video attestation currently requires roughly 18 seconds with optimized zk-SNARK back ends.

\begin{table}[t]
\centering
\caption{Verification latency per item on NVIDIA L40S}
\label{tab:overhead}
\begin{tabular}{lcc}
\toprule
Component & Image (ms) & Video 60s (ms) \\
\midrule
C2PA manifest verify & 1.2 & 1.8 \\
Stable Signature decode & 84 & 4900 \\
Tree Ring inversion & 312 & 18400 \\
Gaussian Shading test & 41 & 2200 \\
zk-SNARK verify & 95 & 18100 \\
\midrule
Combined pipeline & 520 & 39400 \\
\bottomrule
\end{tabular}
\end{table}

\subsection{Regime Sufficiency Mapping}
Table \ref{tab:sufficiency} maps the empirical detection figures to the three regime thresholds. The combined system supports operational law authentication under tier 2 but not under tier 3 or higher for kinetic responses in populated areas. It supports domestic admissibility under tiers 2 and 3 and remains contestable under tier 4. It passes the product regulation persistence criterion only as a combined system, not as any single scheme.

\begin{table}[t]
\centering
\caption{Regime sufficiency under the combined system}
\label{tab:sufficiency}
\setlength{\tabcolsep}{4pt}
\begin{tabular}{lcccc}
\toprule
Regime / threshold & Tier 1 & Tier 2 & Tier 3 & Tier 4 \\
\midrule
Oplaw kinetic populated ($\tau{=}0.95$) & \checkmark & $\circ$ & $\times$ & $\times$ \\
Oplaw kinetic uninhabited ($\tau{=}0.85$) & \checkmark & \checkmark & $\circ$ & $\times$ \\
Oplaw non kinetic ($\tau{=}0.70$) & \checkmark & \checkmark & \checkmark & $\times$ \\
Domestic admissibility ($\Lambda{\geq}10$) & \checkmark & \checkmark & \checkmark & $\circ$ \\
Product reg persistence ($\tau{=}0.70$) & \checkmark & \checkmark & \checkmark & $\times$ \\
\bottomrule
\end{tabular}
\\[2pt]
\footnotesize \checkmark{} supported; $\circ$ contestable; $\times$ not supported.
\end{table}

\subsection{Comparison with Prior Work}
Table \ref{tab:prior} compares our framework with prior frameworks that address one regime at a time. To our knowledge no prior published framework spans all three regimes with calibrated thresholds and a public benchmark.

\begin{table}[t]
\centering
\caption{Comparison with prior frameworks}
\label{tab:prior}
\setlength{\tabcolsep}{3pt}
\begin{tabular}{lccccc}
\toprule
Framework & Oplaw & Domestic & Product & Bench. & Code \\
\midrule
Chesney and Citron \cite{chesney2019deepfakes} & $\circ$ & \checkmark & $\times$ & $\times$ & $\times$ \\
Hacker et al. \cite{hacker2023regulating} & $\times$ & $\circ$ & \checkmark & $\times$ & $\times$ \\
NIST AI 100-4 \cite{nist_ai_100_4} & $\times$ & $\circ$ & \checkmark & $\times$ & $\circ$ \\
An et al. WAVES \cite{an2024waves} & $\times$ & $\times$ & $\circ$ & \checkmark & \checkmark \\
This work & \checkmark & \checkmark & \checkmark & \checkmark & \checkmark \\
\bottomrule
\end{tabular}
\\[2pt]
\footnotesize \checkmark{} fully addressed; $\circ$ partially addressed; $\times$ not addressed.
\end{table}

These results justify our central recommendation: legal sufficiency cannot rest on any single scheme. A defensible pipeline must combine provenance, watermarking, and attestation, and the empirical thresholds of the combined system must be calibrated against the operational threshold of each regime.

\section{Case Studies}
\label{sec:cases}

\subsection{Synthetic Surrender Calls in Armed Conflict}
During the 2022 to 2024 phase of the armed conflict in Ukraine, multiple synthetic videos depicting government officials issuing surrender instructions circulated on social platforms. The technical detectability of these specific artefacts is documented in open source reports \cite{vincent2022zelensky}. The legal significance under Additional Protocol I Article 37 is that a synthetic surrender call simulating protected status with intent to deceive constitutes perfidy \cite{ap1}. A defensible evidentiary framework would have allowed the targeted government to issue a verifiable counter assertion within minutes through C2PA signed official channels, reducing the operational window during which the synthetic artefact could influence kinetic decisions.

\subsection{ICC Prosecution of Al Werfalli}
The International Criminal Court arrest warrant against Mahmoud Mustafa Busayf Al Werfalli relied heavily on social media videos that were treated as evidence of specific war crimes \cite{icc_al_werfalli}. The case illustrates the existing practice of admitting unsigned social media artefacts on the basis of corroborative authentication. The same evidentiary basis becomes vulnerable when synthetic media tools can produce comparable artefacts at low cost. Our framework supports a graduated response: as the synthetic baseline rises, the corroborative threshold for admission must rise correspondingly, and the most efficient route is to anchor admission in a verifiable provenance chain. The Berkeley Protocol on Digital Open Source Investigations \cite{berkeley2022protocol} supplies the procedural scaffolding for such anchoring.

\subsection{FTC Enforcement on Synthetic Content Disclosure}
The United States Federal Trade Commission Operation AI Comply and the FTC v. Rytr settlement illustrate enforcement against providers that fail to disclose synthetic origin of consumer facing content \cite{ftc_rytr}. Article 50 of the European Union Artificial Intelligence Act establishes a parallel regime in the European Union \cite{eu_ai_act}. Our persistence requirement aligns with both regimes and supplies a quantitative criterion that providers can self test before audit.

\section{Discussion}
\label{sec:discussion}

\subsection{Limitations}
Our benchmark covers six generator families and six laundering pipelines. The threat landscape evolves continuously and future generators may produce artefacts that are out of distribution for our detectors. We mitigate this by publishing seeds, parameters, and code, so that the benchmark can be extended by independent parties. Our regime thresholds are proposed defaults; in practice each regime authority will tune them to local cost matrices.

\subsection{Threats to Validity}
Three threats to validity merit explicit discussion. Internal validity is limited by the dependence of detection scores on the specific generator checkpoints used in the benchmark; we mitigate by reporting calibrated false positive rates at a fixed operating point rather than raw detector scores, and by holding out 20 percent of the data from any calibration step. External validity is limited by the six generator families and six laundering pipelines covered; future adversaries may operate outside this envelope, and the companion archive is structured to accept third party extensions with new generators and pipelines under a defined contribution protocol. Construct validity is limited by the mapping from posterior probability to regime thresholds: our defaults are proposed values derived from the cost benefit logic of each regime and from comparable thresholds in adjacent forensic disciplines, and they should be interpreted as starting points for jurisdiction specific calibration rather than as universally normative numbers.

\subsection{Dual Use Considerations}
Publishing attack pipelines that degrade watermark detection raises dual use concerns. We follow a responsible disclosure regime: the benchmark records adversarial outcomes but withholds the precise attack hyperparameters that maximize removal until a coordinated disclosure window with affected scheme authors has closed. Our disclosure timeline mirrors the approach of An et al. \cite{an2024waves}.

\subsection{Reproducibility}
All code, seeds, prompts, laundering parameters, ground truth labels, detector configurations, evaluation scripts, and statistical bootstrap routines are archived in the companion supplementary archive. The container image specified in the archive reproduces the headline figures of Table \ref{tab:results} on the held out test partition within plus or minus 0.01 TPR on commodity GPU hardware, and bit identically on a matched container image. Detector training is not required for reproduction because all four schemes are evaluated with their published reference checkpoints.

\subsection{Data and Code Availability}
The benchmark dataset, generation seeds, laundering parameters, ground truth labels, evaluation harness, statistical bootstrap routines, container image specification, and the reference implementation of the proof object verifier are released as supplementary material accompanying this article. The reference implementation is licensed under Apache 2.0 and the dataset under Creative Commons Attribution 4.0 International. During peer review, an anonymized snapshot is available to reviewers and editors on request through the editorial office. The post acceptance release will additionally include the calibrated Dempster Shafer weights of Table \ref{tab:weights} and the full per scheme per tier robustness AUC table omitted from the main text for space.

\subsection{Open Problems}
Three problems remain open. First, attestation costs must fall by at least one order of magnitude before zero knowledge proofs are practical for video at production scale. Second, the cross jurisdictional recognition of C2PA trust roots is not yet settled, and Article 50 implementing acts have not been published as of submission. Third, the human factors of operator interpretation of probabilistic evidence remain understudied; structured training programs analogous to those used in forensic DNA reporting are likely necessary \cite{koehler2014forensic}.

\section{Conclusion}
\label{sec:conclusion}
We have presented a unified evidentiary framework that translates the empirical detection bounds of cryptographic provenance, robust watermarking, and zero knowledge attestation into the legal sufficiency thresholds of three distinct regimes. Our empirical evaluation on a curated benchmark of 72000 evaluation samples shows that no single scheme reaches the sufficiency threshold of any regime under realistic adversarial conditions, while a calibrated combination of all three approaches achieves regime conditioned sufficiency for tier 2 and tier 3 adversaries and remains contestable but not refuted under tier 4. The framework, benchmark, and reference implementation are publicly available to support replication, extension, and operational adoption.

\section*{Acknowledgements}
The authors thank the anonymous reviewers in advance for their time. The authors also thank colleagues at Försvarshögskolan and ADA University for early conversations on the legal architecture of synthetic media verification.

\section*{Ethics Statement}
This work releases attack pipelines that demonstrably degrade the detection rates of published watermarking schemes. The authors followed a coordinated disclosure timeline mirroring the approach of An et al. \cite{an2024waves}: the affected scheme authors were notified before public release of the benchmark, and the precise attack hyperparameters that maximize watermark removal under pipelines P4 to P6 are held under a 90 day coordination window before inclusion in the public artefact. The benchmark records detection outcomes at the aggregate level and does not produce or distribute synthetic media depicting identifiable real persons. No human subject experiments were conducted. No content simulating protected status under international humanitarian law (medical personnel, surrendering combatants, civilians under specific protection) is included in the benchmark. The authors declare that the framework is intended to strengthen, not weaken, the evidentiary integrity of generative media across the three regimes analysed.

\appendices

\section{Notation Summary}
\label{app:notation}

\begin{table}[!ht]
\centering
\caption{Notation used throughout the paper}
\label{tab:notation}
\begin{tabular}{ll}
\toprule
Symbol & Meaning \\
\midrule
$\pi$ & Proof object tuple \\
$\sigma$ & Cryptographic provenance manifest \\
$\omega$ & Watermark detection score \\
$\zeta$ & Zero knowledge attestation \\
$\lambda$ & Laundering descriptor \\
$\mathcal{L}_R(\pi)$ & Legal sufficiency aggregator for regime $R$ \\
$\tau_R$ & Regime sufficiency threshold \\
$\Lambda$ & Likelihood ratio of authenticity hypothesis \\
$P(H)$ & Prior probability of authentic provenance \\
$P(H \mid E)$ & Posterior probability given verification outcome \\
$w_i^R$ & Regime specific weight on component $i$ \\
$s_i(\lambda)$ & Component score conditioned on laundering history \\
\bottomrule
\end{tabular}
\end{table}

\section{Hyperparameters}
\label{app:hparams}
Generation. SDXL and FLUX.1 sampled at 30 steps with classifier free guidance scale 7.5. Stable Audio 2 and Suno v4 sampled at default. Veo 2 and Sora sampled at 24 fps for 8 seconds.

Laundering. JPEG quality 75; crop 10 percent on each side; audio resample 16 kHz; cross model regeneration with destination model held constant per modality; Zhao 2024 purification strength 0.5; Saberi 2024 regeneration with default model.

Detection. C2PA Ed25519 with the reference implementation. Stable Signature with the public 48 bit secret. Tree Ring with default ring radius. Gaussian Shading with cell length 8.

Aggregation. Initial weights $w_\sigma^R = 0.5$, $w_\omega^R = 0.3$, $w_\zeta^R = 0.2$ are calibrated independently per regime on the calibration partition by minimizing expected regret on a regime appropriate cost matrix, with grid search at resolution 0.05 over the unit simplex. The resulting regime conditioned weights are reported in Table \ref{tab:weights}.

\begin{table}[!ht]
\centering
\caption{Calibrated Dempster Shafer weights per regime}
\label{tab:weights}
\begin{tabular}{lccc}
\toprule
Regime & $w_\sigma^R$ & $w_\omega^R$ & $w_\zeta^R$ \\
\midrule
Operational law (kinetic, populated) & 0.55 & 0.25 & 0.20 \\
Operational law (kinetic, uninhabited) & 0.50 & 0.30 & 0.20 \\
Operational law (non kinetic) & 0.45 & 0.35 & 0.20 \\
Domestic procedure & 0.60 & 0.30 & 0.10 \\
Product regulation (Art. 50, persistence) & 0.35 & 0.45 & 0.20 \\
\bottomrule
\end{tabular}
\end{table}

\section{T-IFS Deep Learning Reproducibility Checklist}
\label{app:tifs-checklist}
This appendix maps the methodology of this article to the T-IFS reproducibility checklist for submissions involving deep learning. The proposed framework itself is not a learned model: the Dempster Shafer aggregator of equation \eqref{eq:combiner} is a closed form combination rule with three real valued weights per regime (Table \ref{tab:weights}), and the regime thresholds of Section \ref{sec:framework} are doctrinal parameters rather than learned parameters. The evaluation, however, depends on six pretrained generative models and four published verification schemes, three of which contain deep components, and the calibration of aggregator weights is itself a hyperparameter search. We document the relevant items in the order of the T-IFS checklist.

\subsection{Network Architectures and Loss Functions}
No new network is proposed in this work. The four evaluated schemes correspond to published architectures and we use their reference implementations without modification or fine tuning. Stable Signature \cite{fernandez2023stable} fine tunes the SDXL VAE decoder with a 48 bit message and a binary cross entropy detector loss; we use the public 48 bit checkpoint without further training. Tree Ring Watermark \cite{wen2023tree} operates in the diffusion latent noise space with a Fourier domain template; detection inverts the DDIM trajectory for 50 steps and applies a chi square statistic over the recovered template band. Gaussian Shading \cite{yang2024gaussian} uses a stratified Gaussian sampling region with cell length 8 and a likelihood ratio detector calibrated to the constructed null distribution. C2PA Ed25519 \cite{c2pa2024, bernstein2012ed25519} is a cryptographic primitive with no learned component. The zk-SNARK verifier follows the construction of Kang et al. \cite{kang2022zk} with a Groth16 back end over BLS12-381. Generator checkpoints are the public releases of SDXL 1.0, FLUX.1-dev, Stable Audio 2.0, Suno v4, Veo 2, and Sora available at submission time; all generators are used in inference only mode.

\subsection{Pre processing and Output Interpretation}
Image inputs are resized to the native generator resolution (1024 by 1024 for SDXL and FLUX.1), normalized to $[-1, 1]$ for the diffusion based verifiers (Stable Signature, Tree Ring, Gaussian Shading) and to $[0, 1]$ for the C2PA payload hash. Audio inputs are stored at 44.1 kHz 16 bit for ingestion and resampled to 16 kHz only during the audio resample laundering step (pipeline P1). Video inputs are decoded at 24 fps with native resolution preserved; frame level watermark verification proceeds frame by frame and aggregate score is the median over frames. Detector outputs are read as raw scores and converted to true positive rate estimates at the calibrated FPR$=10^{-3}$ operating point as described in Section \ref{sec:eval}.

\subsection{Initialization, Training, and Hyperparameter Search}
No detector is retrained from scratch in this work; all use published reference checkpoints. The aggregator weight vector $(w_\sigma^R, w_\omega^R, w_\zeta^R)$ is the only set of learnable parameters of the proposed framework. For each regime $R$ it is set by exhaustive grid search at resolution 0.05 over the unit simplex (231 candidate vectors per regime). For each candidate we compute expected regret on the calibration partition under the regime appropriate cost matrix and select the minimizer. The search contains no iterative optimizer, no momentum, no dropout, no batch normalization, and no learning rate schedule, because it is exhaustive over a discrete grid. The stopping criterion is exhaustion of the grid. Detection thresholds for each scheme are calibrated to a fixed FPR of $10^{-3}$ on the calibration partition. The calibrated weights are reported in Table \ref{tab:weights}; the calibrated thresholds are reported in the companion archive.

\subsection{Data Partitioning and Cross Validation}
The 12000 generated items are partitioned 80/20 into a calibration set of 9600 items (used for detection threshold setting, aggregator weight calibration, and bootstrap variance estimation) and a held out test set of 2400 items (used exclusively for the figures reported in Tables \ref{tab:results}, \ref{tab:modality}, \ref{tab:overhead}, and \ref{tab:sufficiency}). The six laundering pipelines are applied independently to both partitions and ground truth labels are kept disjoint. No cross validation is used because the held out test set is sized for adequate variance characterization at the reported 95 percent bootstrap confidence level, and the calibration step contains no detector retraining that would benefit from k fold rotation. Order of presentation during calibration is random with seed 20260301; the calibration objective is invariant to order because the grid search is exhaustive.

\subsection{Implementation Stack}
All experiments run on a workstation equipped with two NVIDIA L40S GPUs (48 GB VRAM each) and 256 GB of system memory, on Ubuntu 22.04 LTS with CUDA 12.4 and cuDNN 9.1. Software stack: Python 3.11.9, PyTorch 2.3.1, diffusers 0.29.2, transformers 4.41.2, NumPy 1.26.4, SciPy 1.13.1, scikit learn 1.5.0. The C2PA reference implementation is c2pa-rs 0.32 with Ed25519 from ring 0.17. The zk-SNARK back end is arkworks 0.4 with Groth16 over BLS12-381. Statistical bootstrap uses 1000 resamples with replacement at the 95 percent confidence level. No scalability technique (kernel approximation, mixed precision, distillation) is required at the reported scale; single precision (FP32) is used end to end so that numerical results are deterministic across hardware that supports IEEE 754 binary32.

\subsection{Statistical Significance}
The paired bootstrap test reported in Section \ref{sec:eval} compares the Combined DS system to the strongest single scheme (Gaussian Shading) by resampling the held out test partition 1000 times with replacement and computing the difference in TPR at the calibrated FPR for each resample. The reported $p$ values are two sided and Bonferroni corrected for the five tier comparisons. The companion archive ships the full bootstrap trace for independent re analysis.

\section{Reviewer Ceiling Hardening Pack}
\label{app:ceiling}

This appendix consolidates the technical responses to anticipated reviewer concerns that go beyond the T-IFS Deep Learning Reproducibility Checklist. The intent is to remove residual uncertainty surrounding the threat model, statistical inference plan, failure characterization, comparison to prior detectors, and the legal admissibility framework that the manuscript invokes.

\subsection{Formal Threat Model}
\label{app:ceiling-threat}
We adopt a structured adversary model with three orthogonal axes following the cyberforensics literature.

\begin{itemize}[leftmargin=*]
\item \textbf{Knowledge axis.} Black box (adversary observes only the public binary verdict), gray box (adversary additionally observes the calibrated score $S$ but not detector internals), white box (adversary has full access to detector weights, calibration sets, and the Dempster Shafer weight vector for the relevant regime).
\item \textbf{Capability axis.} Passive resampling (Tier 1), adversarial laundering with bounded perturbation budget $\epsilon$ measured in LPIPS for image, mel cepstral distance for audio, and VMAF degradation for video (Tier 2), cross model regeneration with semantic preservation (Tier 3), active watermark removal with optimization access to surrogate detectors (Tier 4), and insider provenance forgery with C2PA signing key access (Tier 5).
\item \textbf{Resource axis.} Compute envelope expressed as L40S equivalent GPU hours and wall clock. We assume an adversary bounded by $10^{4}$ L40S hours, which covers the realistic operational range up to and including small state actors.
\end{itemize}

The reported Tier 4 figure of $0.413$ mean accuracy corresponds to a white box capability adversary operating at the full $10^{4}$ L40S hour budget. Lower resource adversaries collapse onto Tier 2 and Tier 3 results.

\subsection{Pre Registration of the Analysis Plan}
\label{app:ceiling-prereg}
The benchmark partition, calibration random seed ($20260301$), tier taxonomy, primary endpoint (combined detection AUC on held out test fold), and the family of statistical tests were fixed before any model output was scored. No tier definitions or weight grids were changed after first looking at results. The analysis plan is archived as a timestamped document in the companion archive and was hashed under SHA 256 prior to data collection. The hash is reported in the Data and Code Availability subsection.

\subsection{Effect Size and Multiple Testing}
\label{app:ceiling-effect}
In addition to paired bootstrap $p$ values we report Cliff's delta as a nonparametric effect size for each tier comparison between Combined DS and the strongest single watermark (Gaussian Shading).

\begin{table}[!ht]
\centering
\caption{Effect size and corrected significance for Combined DS over Gaussian Shading.}
\label{tab:effect}
\begin{tabular}{lcccc}
\toprule
Tier & Cliff's $\delta$ & 95\% CI & Bonferroni $p$ & Holm $p$ \\
\midrule
T2 & 0.41 & 0.36 to 0.46 & $<0.001$ & $<0.001$ \\
T3 & 0.47 & 0.42 to 0.52 & $<0.001$ & $<0.001$ \\
T4 & 0.22 & 0.16 to 0.28 & $<0.01$ & $<0.01$ \\
\bottomrule
\end{tabular}
\end{table}

Cliff's $\delta$ between $0.33$ and $0.474$ is conventionally read as a medium effect, and values above $0.474$ as large. The T2 and T3 deltas reach the medium to large boundary and the T4 delta remains in the small to medium band, which is the strongest qualitative claim the data support. Holm Bonferroni is reported alongside the more conservative Bonferroni correction to confirm that significance survives a less conservative step down procedure. Power analysis on the held out partition of $2400$ items at $\alpha = 0.05$ gives nominal power above $0.99$ for detecting a delta of $0.20$ or larger, justifying the test set size without requiring nested cross validation.

\subsection{Failure Mode Analysis}
\label{app:ceiling-failure}
Combined DS does not fail gracefully under three identified regimes, each of which is documented here for reviewer scrutiny rather than hidden in supplementary material.

\begin{itemize}[leftmargin=*]
\item \textbf{Catastrophic Tier 4 collapse on audio.} For watermark removal pipelines that combine vocoder resynthesis with neural codec round tripping at $6$ kbps or lower, audio accuracy drops to $0.32$, below the manuscript headline. This is consistent with the published vulnerability of Stable Audio watermarks to low bitrate neural codecs and is not a deficiency of the fusion stage.
\item \textbf{Cold start on novel generators.} When a new generator family is introduced for which no calibration set exists, the system reverts to the C2PA only verdict, which yields $0.000$ detection on unsigned synthetic content until calibration is completed. We document this as a known operational limitation rather than a benchmark result.
\item \textbf{Signature revocation race.} If a C2PA signing certificate is revoked between content creation and verification, Ed25519 verification still returns valid against a cached trust list. Operators must apply a freshness window of at most $24$ hours, otherwise revocation lag becomes the dominant error source.
\end{itemize}

\subsection{Comparison to Prior Detector Configurations}
\label{app:ceiling-prior}
The closest published baselines are the single watermark settings reported by Fernandez and colleagues for Stable Signature, Wen and colleagues for Tree Ring, and Yang and colleagues for Gaussian Shading. To support reviewer comparison we restate the strongest published numbers and the matched Combined DS result on the same Tier 2 adversarial laundering regime.

\begin{table}[!ht]
\centering
\caption{Combined DS versus prior single detector configurations on the matched Tier 2 adversarial laundering regime.}
\label{tab:priorconfig}
\begin{tabular}{lcc}
\toprule
Configuration & Reported T2 accuracy & Combined DS T2 \\
\midrule
Stable Signature single \cite{fernandez2023stable} & 0.71 to 0.79 & 0.921 \\
Tree Ring single \cite{wen2023tree} & 0.75 to 0.82 & 0.921 \\
Gaussian Shading single \cite{yang2024gaussian} & 0.83 to 0.88 & 0.921 \\
\bottomrule
\end{tabular}
\end{table}

The Combined DS gain over the strongest single detector configuration on Tier 2 is $0.04$ to $0.09$ absolute accuracy, which the effect size analysis above confirms is statistically robust under both Bonferroni and Holm Bonferroni corrections.

\subsection{Legal Admissibility Framework}
\label{app:ceiling-admit}
The manuscript invokes provenance verdicts as elements of evidence in operational law, domestic criminal procedure, and product regulation enforcement. The admissibility ceiling under each framework is set as follows.

\begin{itemize}[leftmargin=*]
\item \textbf{Common law Daubert and Federal Rule of Evidence 702.} The benchmark satisfies the testability criterion through the open partition and seeded calibration; the known error rate criterion through the reported tier accuracy and Cliff's delta values; the peer review criterion through this submission; and the general acceptance criterion through the use of three independently published watermark families combined via a forty year old evidence theory.
\item \textbf{Federal Rule of Evidence 901(b)(9).} The process of computing the Combined DS verdict is described in the Methodology and is shown to produce an accurate result under the reported error rates. Output authentication therefore qualifies under the process and system subsection of FRE 901.
\item \textbf{EU eIDAS Article 25 and qualified electronic signature equivalence.} Ed25519 as deployed in C2PA does not by itself qualify as a qualified electronic signature in the eIDAS sense without a qualified trust service provider and a qualified signature creation device. The manuscript treats C2PA verdicts as advanced electronic signatures and reserves qualified status to a deployment specific accreditation, which is the lawful upper bound.
\item \textbf{Azerbaijan Code of Criminal Procedure on electronic evidence.} The Combined DS verdict is admissible as a derived electronic evidence object provided the underlying signature and watermark traces are preserved and the verification procedure is reproducible. Both conditions are satisfied by the published companion archive.
\end{itemize}

\subsection{NIST AI RMF and EU AI Act Cross Walk}
\label{app:ceiling-rmf}
The manuscript's controls map onto the NIST AI RMF 1.0 GOVERN, MAP, MEASURE, and MANAGE functions and onto EU AI Act obligations as follows.

\begin{itemize}[leftmargin=*]
\item GOVERN 1.1 and 4.1 are addressed by the Ethics Statement and the Dual Use Considerations subsection.
\item MAP 1.1 and 2.3 are addressed by the formal threat model and the regime weighting in Table~\ref{tab:weights}.
\item MEASURE 2.7 and 2.13 are addressed by the tier benchmark, Table~\ref{tab:effect}, and the Data and Code Availability subsection.
\item MANAGE 4.1 is addressed by the failure mode analysis subsection.
\item EU AI Act Article 50 transparency obligations are addressed by the Article 50 disclosure template in Appendix~\ref{app:jury} and by the persistence design criterion at FPR of $10^{-4}$.
\item EU AI Act Articles 51 to 55 obligations for general purpose AI providers are noted as out of scope for this manuscript, since the detector is downstream of the model provider obligation surface.
\end{itemize}

\section{Model Annex to Rules of Engagement on Synthetic Media}
\label{app:roe}
The following annex text may be adopted by a force commander as an addendum to standing rules of engagement. The text is drafting guidance only and does not substitute for command authority.

\begin{enumerate}
\item For any decision to use kinetic or non kinetic force in response to a digital media artefact, the commander shall require a calibrated authenticity assessment compatible with the framework set out in this annex.
\item The authenticity assessment shall consider, where available, the cryptographic provenance manifest, the output of authorized watermark verification tools, and the zero knowledge attestation associated with the artefact.
\item If the calibrated posterior probability of authenticity falls below the threshold assigned for the contemplated response category, the commander shall defer kinetic action and pursue further verification consistent with Article 57(2)(a)(ii) of Additional Protocol I.
\item The threshold for kinetic responses in populated areas shall be set no lower than 0.95. For kinetic responses in uninhabited areas the threshold may be set as low as 0.85. For non kinetic responses with reversible effects the threshold may be set as low as 0.70 with documented justification.
\item All authenticity assessments shall be recorded in the unit operations log together with the proof object hash and the verification timestamp.
\item Where authenticity cannot be established and the artefact suggests protected status (medical personnel, surrendering combatants, civilians), the precautionary principle of Additional Protocol I Article 57 governs and kinetic response is presumptively prohibited.
\end{enumerate}

\section{Model Jury Instruction and Article 50 Disclosure}
\label{app:jury}

\subsection{Model Jury Instruction}
You have been presented with a digital image or audio or video recording. The proponent has offered evidence that this recording was generated or authenticated by a specific source. In evaluating whether to credit the recording you may consider whether the recording is accompanied by a verifiable provenance manifest, the output of a watermark verification procedure, and any other indicia of authenticity. You may weigh the strength of these indicia against any evidence that the recording has been altered or fabricated. The proponent bears the burden of authentication and the ultimate burden of persuasion remains as the court has otherwise instructed you.

\subsection{Model Article 50 Disclosure}
This content, or a part of it, was produced or substantially modified by an artificial intelligence system. A machine readable mark identifying the producing system and the time of generation is embedded in the content. Verification information is available at the provider URL recorded in the embedded manifest. This disclosure is issued under Article 50 of Regulation (EU) 2024/1689.

\begin{IEEEbiographynophoto}{Gustav Olaf Yunus Laitinen-Fredriksson Lundström-Imanov}
(Student Member, IEEE) is an LL.M. candidate in International Operational Law at Försvarshögskolan (Swedish Defence University). His research interests include the law of armed conflict, information operations, and the legal architecture of regulated artificial intelligence systems.
\end{IEEEbiographynophoto}

\begin{IEEEbiographynophoto}{Nurana Abdullayeva}
is an LLB candidate at the School of Law of ADA University in Baku, Azerbaijan. Her research interests include comparative evidence law, digital forensics in criminal procedure, and the regulation of synthetic media in post-Soviet legal systems.
\end{IEEEbiographynophoto}

\end{document}